\documentclass[showpacs,floatfix,twocolumn]{revtex4-1}
\usepackage{graphicx,psfrag,amsmath,amssymb,amsfonts,latexsym,color,epsf,graphpap,dcolumn}

\begin{document}

\title{Derivative expansion for the electromagnetic and Neumann Casimir
effects in $2+1$ dimensions with imperfect mirrors}
\author{C.~D.~Fosco$^{a}$, F.~C.~Lombardo$^{b}$, F.~D.~Mazzitelli$^{a}$}
\affiliation{$^a$Centro At\'omico Bariloche and Instituto Balseiro, 
Comisi\'on Nacional de Energ\'\i a At\'omica, R8402AGP Bariloche, Argentina.\\
$^b$Departamento de F\'\i sica {\it Juan Jos\'e
Giambiagi}, FCEyN UBA and IFIBA CONICET-UBA, Facultad de Ciencias Exactas y Naturales,
Ciudad Universitaria, Pabell\' on I, 1428 Buenos Aires, Argentina}

\date{today}

\begin{abstract}
\noindent 
We calculate the Casimir interaction energy in $d=2$ spatial dimensions
between two (zero-width) mirrors, one flat, and the other slightly 
curved, upon which {\em imperfect\/} conductor boundary conditions are imposed for an
Electromagnetic (EM) field. Our main result is a second-order Derivative Expansion (DE) 
approximation for the Casimir energy, which is studied in different interesting limits.  
In particular, we focus on the emergence of a non-analyticity beyond the
leading-order term in the DE, when approaching the limit of
perfectly-conducting mirrors.  
We also show that the system considered is equivalent to a dual one,
consisting of a massless real scalar field satisfying imperfect Neumann
conditions (on the very same boundaries). Therefore, the results obtained
for the EM field hold also true for the scalar field model.
\end{abstract}

\maketitle

\section{Introduction}\label{sec:intro}

The static Casimir force is a physical effect which manifests itself in systems
consisting of a fluctuating (quantum, thermal,\ldots) field in the presence
of non trivial, time-independent boundary conditions \cite{booksCasimir}. 
The corresponding Casimir energy, may be characterized as a real-valued
{\em functional\/} of function(s) which, under a certain
parametrization, define the geometry of the boundaries.  
This observation can be used as the starting point for an approximation
scheme, the Derivative Expansion (DE)
originally proposed in Ref.\cite{de1} (for subsequent developments see
Ref.\cite{de2}).
The DE adopts its simplest form when
the boundary conditions considered are `perfect', i.e., they do not involve
any parameter, and the geometry of the system is sufficiently simple, yet
non trivial: One has two boundaries, one of them, $R$, is smoothly curved,
and describable in terms of a single-valued `height function' $\psi$, which
measures the vertical distance of each one of its points to the flat boundary, $L$.
In other words, such that $R$ can be projected in terms
of a single Monge patch, with $L$ the  projection plane.  

Under the assumptions above, one is clearly left with an energy which is a
functional of a single function, $\psi$, and also possibly a function of the
parameters eventually appearing in the definition of the boundary
conditions (specially when they are imperfect). The DE is an approximation scheme for
that functional,  such that its leading-order term is tantamount to the
proximity force approximation (PFA) \cite{pfa}. 

The nature of the next-to-leading-order (NTLO) term, on the other hand,
depends on the type of boundary condition being imposed on the field.  For
Dirichlet boundary conditions, it has been shown that it is always
quadratic in the derivatives of the smooth functions $\psi$, regardless of
the number of spatial dimensions, $d$ \cite{ded}. Therefore its contribution to the
Casimir energy is the integral of a local function.  The same happens for
perfect Neumann conditions when $d \neq 2$, but the situation is qualitatively
different when $d=2$: the NTLO term becomes nonlocal in coordinate space, a
phenomenon which, we have argued, is a manifestation of the existence of a
massless excitation for the fluctuating field \cite{ded}. A similar effect may be seen
to appear for the case of the EM field with perfect boundary conditions,
also in $d=2$. This is, as we shall show below, no coincidence, as both
theories, scalar with Neumann conditions and EM field with perfect boundary
conditions are equivalent.

In order to gain more insight about this issue, namely, the special nature
of the NTLO contribution for the EM field in $2+1$ spacetime dimensions
with perfect boundary conditions, we perform here the following analysis:
we consider a system with imperfect-conductor boundary conditions on both
surfaces, and evaluate the leading and NTLO contributions to the DE. 

This study is of interest because of several reasons: on the one hand, one
knows that the perfect-conductor condition is an approximation to a real,
imperfect mirror. Besides, it will provide a way to cope with the infrared
divergences which would appear for perfect conditions.  Finally, note that,
in spite of the fact that the system is defined in $2+1$ dimensions, this
analysis may be useful even when one considers the $3+1$-dimensional case at
finite temperature. Indeed, thermal effects mean that one should take into
account the contribution of the Matsubara modes~\cite{finiteT}. Among them, the thermal
zero mode behaves as a $d=2$ field and, as we have shown, an entirely
analogous effect to the one has in $d=2$ is induced~\cite{ded}. 
The same can be said of a fluctuating Electromagnetic (EM) field with
perfect boundary conditions at a finite temperature, since it can be shown to accommodate a
Neumann like contribution.

This paper is organized as follows: in Sect.~\ref{sec:thesys} we define the
system, starting by a  description of the duality between the scalar and EM
field models, and then constructing the model for the EM field coupled to the two
boundaries. We then define the respective effective action, functional of
the shape of the deformed mirror.
In Sect.~\ref{sec:exp} we present results about the expansion of the
effective action to second order in the departure with respect to the case
of flat parallel mirrors.  Then, in Sect.~\ref{sec:de} we deal with the DE
for the Casimir energy, based on the results obtained in the previous
Section.  In Sect. \ref{sec:examp} we present some examples where 
the NTLO correction is evaluated for  
imperfect Neumann boundary conditions in 2+1 dimensions.  
Sect.~\ref{sec:conc} contains our conclusions.
Some technical details about the evaluation of the effective action to the
second order in the deformation are presented in an Appendix.
\section{The system and its effective action}\label{sec:thesys}
\subsection{Scalar field / EM field duality}
 Let us first see how a real scalar field in $2+1$ dimensions with Neumann
conditions may be described, alternatively, in terms of an EM field with
conductor boundary conditions. 

We first assume that we want to study a massless quantum real scalar field
$\varphi(x)$, satisfying {\em perfect\/} Neumann boundary conditions on
two static curves, denoted by $L$ and $R$, the former assumed to be
straight and the latter slightly curved.  We use Euclidean conventions,
such that $x=(x_\mu)$, with \mbox{$\mu=0,1,2$}, denote the
$(2+1)$-dimensional spacetime coordinates ($x_0 \equiv$ imaginary time).
Besides, we shall use the notation~\mbox{$v_\parallel$} for a vector
on a $1+1$ dimensional spacetime, e.g.  \mbox{$v_\parallel
= (v_\alpha)=(v_0,v_1)$}. We have introduced the convention, that we
follow in the rest of this paper, that indices from the beginning of the
Greek alphabet ($\alpha, \,\beta,\ldots$) run over the values $0$ and $1$. 
No distinction will me made between upper and lower indices, and their
vertical position will only be decided having notational clarity in mind. 

The free Euclidean action for the vacuum field $\varphi$ is given by
\begin{equation}\label{eq:defsphi}
{\mathcal S}_0\;=\; \frac{1}{2}\,\int d^3x \, (\partial \varphi)^2\;,
\end{equation}
which is complemented by the assumption of Neumann boundary
conditions on L and R. Regarding the Casimir energy
calculation we just need static boundaries, but it is nevertheless useful
to consider a more general, time-dependent expression for the curved
boundary R. Thus, L and R are defined as the regions in spacetime
satisfying the equations:
\begin{eqnarray}
	{\rm L}) \;\;\; x_2 &=& 0 \;\;,\nonumber\\
	{\rm R}) \;\;\; x_2 &=& \psi(x_\parallel)\;\;,
\end{eqnarray}
respectively.

The reason to allow for such a time dependence is twofold; on the one hand the
treatment of the problem is more symmetrical, and the physical case may
still be recovered at the end by setting
$\psi=\psi(x_1)$. On the other hand, the Euclidean effective action which
we shall calculate can be used (reinterpreted) as the high temperature
limit of the free energy for a model in $3+1$ dimensions, with boundaries
defined by $x_3 =0$ and $x_3 = \psi(x_1,x_2)$ (after a straightforward
relabelling of the spacetime coordinates).

The form of the boundary conditions imposed on the field at the
boundaries (regarded as spacetime surfaces) is then
\begin{equation}\label{eq:nbcs}
\partial_2\varphi(x)\Big|_{x_2 = 0} = 0 \;\;,\;\;
\partial_n \varphi(x)\Big|_{x_2 = \psi(x_\parallel)} = 0 \;,
\end{equation}
where \mbox{$\partial_{n}$} denotes the directional derivative along the
direction defined by the unit normal to R, $n^\mu(x_\parallel)$: 
\begin{equation}
n^\mu(x_\parallel) \;\equiv\; \frac{ \delta^\mu_2 - \delta^\mu_\alpha \partial_\alpha
\psi(x_\parallel) }{\sqrt{1+|\partial \psi(x_\parallel)|^2}} \;.
\end{equation}

The scalar field $\varphi$ may then be mapped into the $3$-potential
$A_\mu$ for an EM field, by means of the duality transformation:
\begin{equation}\label{eq:duality}
\partial_\mu \varphi \;\leftrightarrow\; \epsilon_{\mu\nu\rho}\partial_\nu A_\rho
\end{equation} 
where $A_\mu$ is a vector field. It is an immediate consistency condition
of the above, by taking the divergence on both sides, that $\varphi$ is
massless, $\Box \varphi = 0$, which we shall assume. 

Now, the boundary conditions (\ref{eq:nbcs}) corresponds, via the duality
transformation, to: 
\begin{eqnarray}
\epsilon_{\alpha\beta} \partial_\alpha A_\beta (x_\parallel, 0) &=& 0
\nonumber\\
\big[ n^\mu(x_\parallel) \, \epsilon_{\mu\nu\rho} \partial_\nu A_\rho
\big]\Big|_{x_2 = \psi(x_\parallel)} &=& 0 \;,
\end{eqnarray}
for the $A_\mu$ field. This may be expressed equivalently as the 
vanishing of the component of the EM field tensor which is `parallel' to the
respective surface; namely, the component of its dual (a pseudo-vector)
parallel to the normal at each point vanishes.  
Let us consider that for R, since the situation for
L may be obtained as a particular case, namely, $\psi =0$:
For R, introducing the projected component of the gauge field,
${\mathcal A}_\alpha(x_\parallel)$: 
\begin{eqnarray}
{\mathcal A}_\alpha(x_\parallel) \, &\equiv &\, e^\mu_\alpha(x_\parallel)
A_\mu(x_\parallel,\psi(x_\parallel)) \nonumber \\
e^\mu_\alpha(x_\parallel) \; &\equiv &\; \delta^\mu_\alpha + \delta^\mu_3
\partial_\alpha \psi(x_\parallel)  \;,
\end{eqnarray}
one gets on the surface the boundary condition:
\begin{equation}\label{eq:condbcs}
\epsilon_{\alpha\beta}\partial_\alpha {\mathcal
A}_\beta(x_\parallel)  \;=\; 0 \;,
\end{equation}
which for $L$ simplifies to:
\begin{equation}\label{eq:condbcsl}
\epsilon_{\alpha\beta}\partial_\alpha A_\beta(x_\parallel,0)  \;=\; 0 \;,
\end{equation}
i.e., the $E_1$ component of the electric field vanishes for $x_2=0$.

So, regarding the boundary conditions, we have a mapping between 
 Neumann and perfect conductor; for the respective free Euclidean
Lagrangians, we note that, from (\ref{eq:duality}):
\begin{equation}
\frac{1}{2} \partial_\mu \varphi \partial_\mu \varphi \;=\; \frac{1}{4} F_{\mu\nu}F_{\mu\nu}
\end{equation}
so that the free scalar field action is mapped into the Maxwell action:
\begin{equation}
\frac{1}{2} \int d^3x  (\partial \varphi)^2  \,=\, {\mathcal S}_0[\varphi]
\;\leftrightarrow \;{\mathcal S}_0[A] \,=\, \frac{1}{4} \int d^3x F_{\mu\nu}^2(A) \;.
\end{equation}
Now we want to deal with {\em approximate\/} boundary conditions, of such a
kind that, for the scalar field would correspond to adding to the action an
 interaction term ${\mathcal S}_I$ localized on two mirrors, and containing
the parameter $\mu$, with the dimensions of a mass, such that perfect
conditions are recovered when $\mu \to 0$.
More explicitly,
\begin{eqnarray}\label{intscalar}
{\mathcal S}_{\rm I}[\varphi] &=& \frac{1}{2\mu} \int d^3x \left[ 
\delta(x_2) (\partial_2\varphi(x))^2 \right.\nonumber \\
&+& \left. \sqrt{g(x_\parallel)} \delta(x_2 - \psi(x_\parallel)) (\partial_n
\varphi(x))^2 \right] \;,
\end{eqnarray}
where $g(x_\parallel) = 1 + [\partial \psi(x_\parallel)]^2$ is the
determinant of the induced metric on R, required to have
reparametrization invariance. We use the same $\mu$ on both mirrors, since
we will assume them to have identical properties, differing just in their
position and geometry.

\subsection{The EM field model}

The approximate Neumann boundary conditions are then introduced in terms of
the EM field, by adding to the EM field action the respective interaction
term. Thus, we will work with the action:
\begin{equation}
{\mathcal S}(A) \;=\; {\mathcal S}_0(A) \,+\, {\mathcal S}_{\rm I}(A) 
\end{equation}
where
\begin{equation}
{\mathcal S}_0(A) \;=\; \frac{1}{4} \int d^3x F_{\mu\nu}F_{\mu\nu}\;,
\end{equation}
and
\begin{equation}
{\mathcal S}_{\rm I}(A) \;=\; {\mathcal S}_{\rm L}(A) \,+\, {\mathcal S}_{\rm R}(A) \;. 
\end{equation}
where
\begin{equation}
{\mathcal S}_{\rm R}(A) \;=\; \frac{1}{4\mu} \int d^2x_\parallel \,\sqrt{g(x_\parallel)} 
{\mathcal F}_{\alpha\beta}(x_\parallel) {\mathcal F}_{\alpha\beta}(x_\parallel)
\end{equation}
with ${\mathcal F}_{\alpha\beta}(x_\parallel)$ the EM field associated to ${\mathcal
A}_\alpha(x_\parallel)$, and $\mu$ the parameter introduced for the scalar
field.  The factor ${\mathcal S}_{\rm L}(A)$ is defined by a similar expression, obtained by
setting $\psi \equiv 0$:  
\begin{equation}
{\mathcal S}_{\rm L}(A) \;=\; \frac{1}{4\mu} \int d^2x_\parallel \, 
F_{\alpha\beta}(x_\parallel,0) F_{\alpha\beta}(x_\parallel,0) \;.
\end{equation}
The interaction terms reproduce Eq.(\ref{intscalar}) when written in terms of the dual
scalar field.

Following standard procedures \cite{fosco2012imp}, we rewrite the action in an equivalent form,
by using two auxiliary fields, $\xi_L$ and $\xi_R$, living on each one of the
surfaces, to linearize the form of the terms localized on the mirrors. 
Those auxiliary fields are introduced in such a way that, if integrated
out, they reproduce the original action, ${\mathcal S}[A]$.

The corresponding equivalent action thus becomes:
\begin{equation}
{\mathcal S}(A,\xi_{\rm L}, \xi_{\rm R}) ={\mathcal S}_0(A) + {\mathcal
S}_0(\xi_{\rm L}, \xi_{\rm R})-i \int d^3x J_\mu(x) A_\mu(x)   
\end{equation} 
with
\begin{eqnarray}
J_\mu(x) &=& \delta(x_2)\, \delta_{\mu \alpha} \epsilon_{\alpha\beta}\partial_\beta \xi_{\rm L}(x_\parallel) \nonumber \\ &+&
\delta(x_2-\psi(x_\parallel))\, e^\mu_\alpha(x_\parallel)
\epsilon_{\alpha\beta}\partial_\beta \xi_{\rm R}(x_\parallel) \;,
\end{eqnarray}
and
\begin{equation}
{\mathcal S}_0(\xi_{\rm L}, \xi_{\rm R})\,=\, \frac{\mu}{2}\int d^2x_\parallel
\Big[ \big( \xi_L(x_\parallel) \big)^2 + \sqrt{g(x_\parallel)} \,\big(
\xi_{\rm R}(x_\parallel) \big)^2 \Big]\;,
\end{equation}
with $g(x_\parallel) = 1 + \partial_\alpha\psi \partial_\alpha\psi$.

The action ${\mathcal S}(A,\xi_{\rm L}, \xi_{\rm R})$ may be seen to be invariant under
gauge transformations, $A_\mu(x) \to A_\mu(x) + \partial_\mu \omega(x)$ as
a consequence of the fact that the `current' $J_\mu$ is conserved. Since
our next step amounts to integrating our the $A_\mu$, that action should be
first given a gauge fixing. Gauge invariance assures the results are going
to be independent of the gauge-fixing adopted, thus our choice is dictated
by simplicity. In this case that is the Feynman gauge, whereby one ads a
gauge-fixing action $S_{\rm gf}(A)$ to ${\mathcal S}_0(A)$, to get the free
gauge fixed action ${\mathcal S}_{\rm G}(A)$:
\begin{equation}
{\mathcal S}_0(A) \;\to\; {\mathcal S}_{\rm G}(A) \,=\, 
{\mathcal S}_0(A) \,+\,\frac{1}{2}\int d^3x (\partial \cdot A)^2 \;.
\end{equation} 

Now we define the effective action $\Gamma[\psi]$ by the functional integral:
\begin{equation}
e^{-\Gamma[\psi]}\;=\; 
\frac{\int {\mathcal D}A {\mathcal D}\xi_{\rm L}{\mathcal D}\xi_{\rm R} \;e^{-{\mathcal
S}_{\rm G}(A)+\,i\, \int d^3x J_\mu(x) A_\mu(x)}}{\int {\mathcal D}A \;e^{-{\mathcal S}_{\rm G}(A)}} ,
\end{equation}
where the denominator has been introduced in order to get rid of one
infinite factor which is irrelevant to our calculation: the effective
action corresponding to the EM field in the vacuum, i.e., in the absence of
mirrors. There are other factors we will get rid of in $\Gamma$, associated
to the self-energies of the mirrors. These have the distinctive feature of
being independent of the distance between the mirrors, and therefore they
do not contribute to the Casimir force between them.

We then integrate  $A$, obtaining for $\Gamma[\psi]$ a formal expression
where we have to integrate over the auxiliary field, which are endowed with
an action we denote by ${\mathcal S}_{\rm eff}(\xi_{\rm L}, \xi_{\rm R})$:
\begin{equation}
e^{-\Gamma[\psi]}\;=\; 
\int {\mathcal D}\xi_{\rm L}{\mathcal D}\xi_{\rm R} \;e^{-{\mathcal S}_{\rm eff}(\xi_{\rm L}, \xi_{\rm R})}
\end{equation}
where 
\begin{eqnarray}
{\mathcal S}_{\rm eff}(\xi_{\rm L}, \xi_{\rm R}) &=& \frac{1}{2} \int d^3x d^3x' 
J_\mu(x) G_{\mu\nu}(x,x') J_\mu(x') \\
&=& \frac{1}{2} \int d^2x_\parallel d^3x'_\parallel \; \xi_{\rm A}(x_\parallel)
K_{\rm AB}(x_\parallel,x'_\parallel) \xi_{\rm B}(x'_\parallel)  \nonumber ,
\end{eqnarray}
where
\begin{equation}
G_{\mu\nu}(x,x') \;=\; \delta_{\mu\nu} \int \frac{d^3k}{(2\pi)^3}
\frac{e^{i k \cdot (x-x')}}{k^2} \;,
\end{equation}
is the $A_\mu$ propagator in the Feynman gauge, and we have introduced the
four objects $K_{\rm AB}(x_\parallel,x'_\parallel)$, where A and B adopt
the values L and R. Their form is obtained by
substitution of the explicit form of $J_\mu$ in terms of the auxiliary
fields, performing integrations by parts, and using the respective
$\delta$ functions. Since we will use an expansion in powers of the deformation
$\eta$, we only need them up to the second order in that expansion. 
See the Appendix for their explicit forms.

We conclude this section by writing the effective action as follows:
\begin{equation}
\Gamma[\psi] \;=\; \frac{1}{2} {\rm Tr} \log[ {\mathbb K}] \;,
\end{equation}
where ${\mathbb K}$ is the $2\times 2$ matrix of components defined by the
kernels $K_{\rm AB}$, while the trace operation acts over the A, B indices,
as well as over the spacetime dependencies.

\section{Expansion of $\Gamma[\psi]$ up to second order in
$\eta$}\label{sec:exp}
As we have already done in previous applications of the DE, we consider the
effective action for an, in principle, time-dependent function $\psi$,
taking the static limit at the end of the calculation. In that limit, the
effective action becomes equal to the vacuum energy $E$, times $T$, the
length of the time coordinates.

Setting then $\psi(x_\parallel) \;=\; a \,+\, \eta(x_\parallel)$,
we introduce the expansion of $\Gamma$ in powers of $\eta$, up to the second order. Thus,
\begin{equation}\label{eq:expansion}
\Gamma(a,\eta)\;=\; \Gamma^{(0)}(a) \;+\;\Gamma^{(1)}(a,\eta) \;+\;
\Gamma^{(2)}(a,\eta) \;+\;\ldots 
\end{equation}
where the index denotes the order in $\eta$.
The first order term can be made to vanish by a proper definition of $a$,
and we consider the relevant,  zeroth and second orders in the following
subsections. They are:
\begin{eqnarray}\label{eq:fbnn}
	\Gamma^{(0)} &=& \frac{1}{2} \, {\rm Tr} \big[\log {\mathbb
	K}^{(0)}\big]\nonumber\\
	\Gamma^{(2)} &=&  \Gamma^{(2,1)} \,+\,\Gamma^{(2,2)} \;,
\end{eqnarray}
where:
\begin{eqnarray}\label{eq:fbn2}
	\Gamma^{(2,1)} &=& \frac{1}{2} \, {\rm Tr} \Big[\big({\mathbb
K}^{(0)}\big)^{-1}{\mathbb K}^{(2)}\Big] \nonumber\\
	\Gamma^{(2,2)} &=& - \frac{1}{4}\, {\rm Tr} 
\Big[\big({\mathbb K}^{(0)}\big)^{-1}{\mathbb K}^{(1)}
\big({\mathbb K}^{(0)}\big)^{-1}{\mathbb K}^{(1)}\Big] \;.
\end{eqnarray}

\subsection{Leading order}
The leading order term may be obtained rather straightforwardly, since the
zeroth-order kernel is block-diagonal in momentum space, and the trace
operation is then two-dimensional, the result being:
\begin{equation}
\Gamma^{(0)}\;=\; T L \frac{1}{2} \int \frac{d^2k_\parallel}{(2\pi)^2}
\log\Big[ 1 - r^2(|k_\parallel|) e^{-2 |k_\parallel| a}\Big] \;.
\end{equation}
with
\begin{equation}
r(x) \;\equiv\; \frac{x}{x + 2 \mu}\;,
\end{equation}
while $T$ and $L$ denote the extent of the time and length dimensions of
the system. We have extracted from $\Gamma^{(0)}$ an $a$-independent
contribution.

Thus, the energy density to this order has the form:
\begin{equation}
{\mathcal E}^{(0)}\;=\; \frac{1}{ 4 \pi a^2}  \int_0^\infty d\rho \rho 
\log\Big[ 1 - \big(\frac{\rho}{\rho + 2 \mu a}\big)^2 e^{-2 \rho}\Big] \;.
\end{equation}
This expression is well-defined for any value of $\mu a$; in particular,  in
the two limiting regimes corresponding to semitransparent mirrors, $\mu a >> 1$: 
\begin{equation}
{\mathcal E}^{(0)}\;\simeq\; - \frac{3}{128 \pi} \, \frac{1}{\mu^2 a^4} \;,
\end{equation}
as well as for perfect mirrors,  $\mu a << 1$:
\begin{equation}\label{order0perf}
{\mathcal E}^{(0)}\;\simeq\; - \frac{\zeta(3)}{16\pi} \, \frac{1}{a^2} \;.
\end{equation}
This result corresponds, of course,  to that of perfect Neumann boundary conditions.

\subsection{Next to leading order}
The second order term $\Gamma^{(2)}$, being quadratic in $\eta$, can be
represented in Fourier space as:
\begin{equation}
	\Gamma^{(2)}\,=\, \frac{1}{2} \int \frac{d^2k_\parallel}{(2\pi)^2}
	\,f^{(2)}(k_\parallel) \, |\tilde{\eta}(k_\parallel)|^2
\end{equation}
in terms of the kernel $f^{(2)}$.

Collecting the two contributions to $f^{(2)}$ presented in the Appendix, we
obtain its full form. It may be
represented as the sum of two terms: $f_l$,  which is just
quadratic in momentum and therefore gives rise to a local contribution
to the effective action, plus another one, $f_{nl}$ where the dependence in $k$ is
inside the integrand of an integral over another momentum ($p_\parallel$),  and 
gives a nonlocal contribution to the effective action;
namely,
\begin{equation}\label{eq:resf2}
f^{(2)}(k_\parallel) \;=\; f_l(k_\parallel) \,+\, f_{nl}(k_\parallel) \;. 
\end{equation}
with
\begin{equation}\label{eq:deffq}
f_l(k_\parallel)\;=\; k_\parallel^2 \;\Big[\int \frac{d^2p_\parallel}{(2\pi)^2}
B(|p_\parallel|)\, r^3(|p_\parallel|)\,
\, \frac{\mu}{|p_\parallel|} \Big]
\end{equation}
and 
\begin{eqnarray}\label{eq:deffint}
f_{nl}(k_\parallel) &=& - \, \int \frac{d^2p_\parallel}{(2\pi)^2} e^{ 2
|p_\parallel+k_\parallel| a}
\left\{ r^3(|p_\parallel|) \right.  \nonumber\\ &+& \left. r(|p_\parallel|) 
r(|p_\parallel + k_\parallel|)
+ r^2(|p_\parallel|) r^2(|p_\parallel + k_\parallel|) \right. \nonumber \\ 
&-& \left. r^3(|p_\parallel|) r^2(|p_\parallel + k_\parallel|) 
e^{-2 |p_\parallel+k_\parallel|a} \right\} \nonumber\\
&\times& B(|p_\parallel|)B(|p_\parallel+ k_\parallel|)
\frac{[p_\parallel \cdot (p_\parallel + k_\parallel)]^2}{|p_\parallel|
|p_\parallel + k_\parallel|} \;, 
\end{eqnarray}
where we have introduced the function:
\begin{equation}
B(x)= \frac{1}{e^{2 a x} - r^2(x)} \;.
\end{equation}

It is quite straightforward to check that the previous expressions render
the proper limit for the perfect-conductor case, $\mu \to 0$, under which
$r \to 1$: 
$$
f^{(2)}(k_\parallel) \; \to \; f^{(2)}_{\rm N}(k_\parallel)
$$
\begin{eqnarray}
f^{(2)}_{\rm N}(k_\parallel) &=& - 2 \, \int \frac{d^2p_\parallel}{(2\pi)^2}
\frac{1}{(e^{2 |p_\parallel| a}- 1)(1 - e^{- 2 |p_\parallel+k_\parallel|
a})}\nonumber \\
&\times &\frac{[p_\parallel \cdot (p_\parallel + k_\parallel)]^2}{|p_\parallel|
|p_\parallel + k_\parallel|} \;,
\end{eqnarray}
where $f^{(2)}_{\rm N}$ also equals the kernel for a real scalar field with Neumann
boundary conditions \cite{ded}.

\section{Derivative Expansion}\label{sec:de}
The Casimir energy $E$, we have argued, is a functional of $\psi$.  Up to
the second order, and recalling that $\psi$ depends on just one coordinate
($x_1$),  $E_{\rm DE}$, has the form: 
\begin{equation}\label{eq:de2}
E_{\rm DE}[\psi] \;=\; \int_{-\infty}^\infty dx_1 \, 
\Big[ 
V(\psi(x_1)) \,+\, Z(\psi(x_1))   \left(\frac{d\psi(x_1)}{dx_1} \right)^2  \Big] \;,
\end{equation} 
where $V$ and $Z$ are local functions of $\psi$ and
$\mu$.
Taking into account the dimensions of the objects involved, we can write a
more explicit form for $V$ and $Z$:
\begin{equation}
V \;=\; \frac{c_0\big(\mu\psi)}{[\psi(x_1)]^2}\;\;\; 
Z \;=\; \frac{c_2\big(\mu\psi)}{[\psi(x_1)]^2} \;.
\end{equation}
where $c_0$ and $c_2$, which determine the zeroth and second order terms,
respectively, are dimensionless functions of their (also dimensionless)
arguments. 

\begin{figure}
\centering
\includegraphics[width=8cm, angle=0]{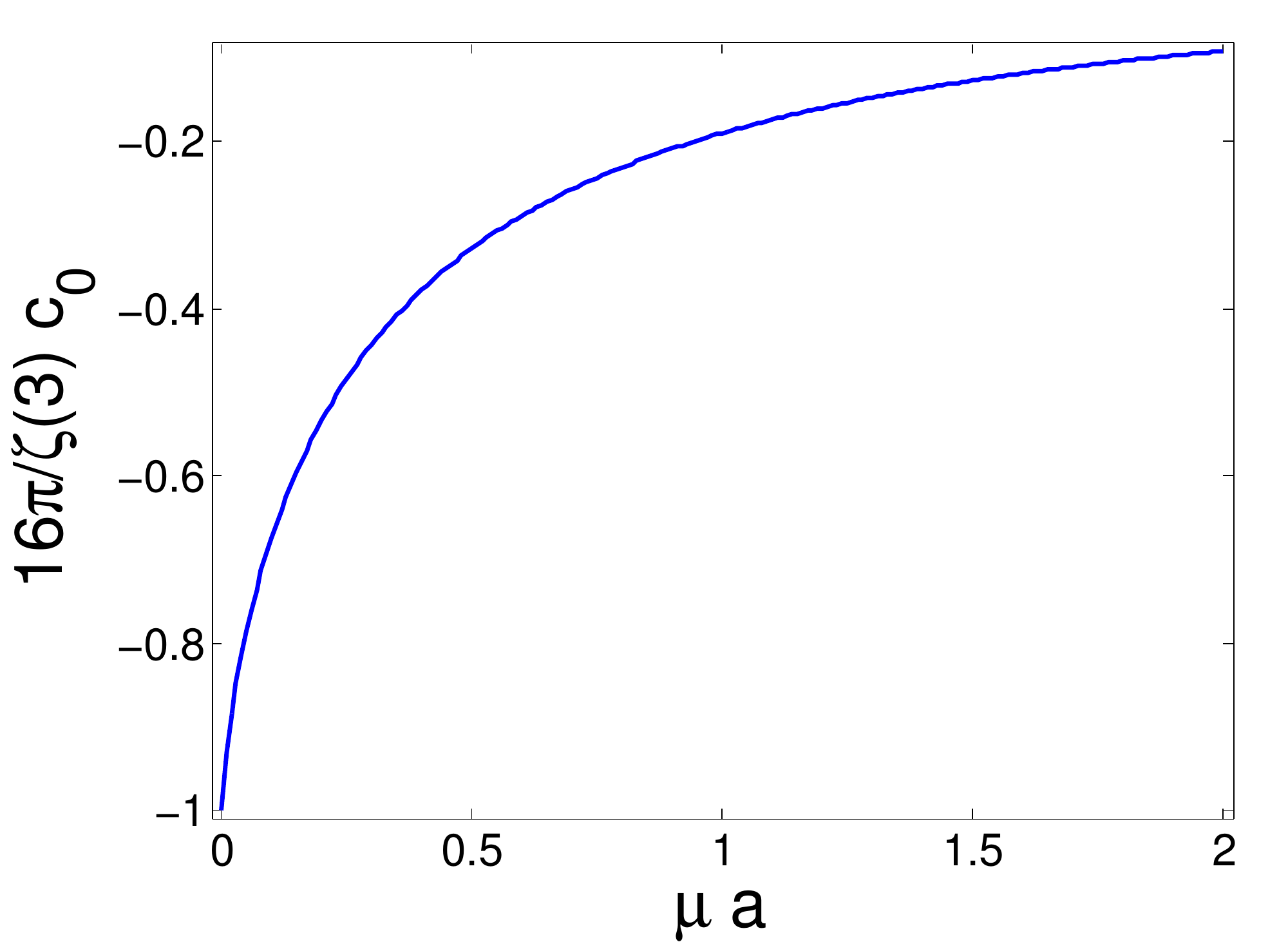}
\caption{Coefficient $16 \pi c_0/\zeta(3)$ as a function of $\mu a$.} \label{fig1}
\end{figure}

Regarding the $c_0$ coefficient, we find that:
\begin{equation}
c_0(\mu a)\;=\; \frac{1}{ 4 \pi}  \int_0^\infty dx \, x
\log\Big[ 1 - \big(\frac{x}{x + \mu a}\big)^2 e^{-2 x}\Big] \;.
\end{equation}
Note that, as shown in Fig. 1,  $c_0(\mu a)$ interpolates between zero in the limit of large $\mu a$, and  
$-\zeta(3)/(16\pi)$ for $\mu a\to 0$, the result that corresponds to perfect Neumann boundary conditions
(see Eq.(\ref{order0perf})).

On the other hand, the $c_2$ coefficient can be extracted from the
kernel appearing in the second order term for the effective action, $\Gamma^{(2)}$, as follows:
In the Taylor expansion around zero-momentum, we denote by $\alpha$
the coefficient of term quadratic in the momentum: 
\begin{equation}\label{eq:e2}
f^{(2)}(k_\parallel) \;=\; f^{(2)}(0) \,+\; \alpha \, k_\parallel^2
\,+\, \ldots
\end{equation}
Then, the $c_2$ coefficient is given by:
$c_2(\mu a) \;=\;  \frac{a^2}{2} \, \alpha \;.$

The contribution to $c_2$ coming from $f_l$ can be
obtained from Eq.(\ref{eq:deffq}). Performing the angular integration, it reads
\begin{equation}
c_{2l}(\mu a)= \frac{\mu a}{4 \pi}\int_0^\infty dx \, \frac{x^3}{(x+2\mu a)}\frac{1}{(x+2\mu a)^2e^{2x}-x^2}     \, .
\end{equation}

The calculation of the other contribution to $c_2$, coming from $f_{nl}$,
is lengthy but straightforward. We expand the integrand in
Eq.(\ref{eq:deffint}) in powers of $k_\parallel$ keeping just quadratic terms. Performing the angular integration we obtain
\begin{eqnarray}\label{c2nl}
&&c_{2nl}(\mu a)=\frac{-1} {4 \pi}\int_0^\infty dx\frac{1}{(2 \mu a+x) \left(x^2-e^{2 x} (2 \mu a+x)^2\right)^4}\nonumber\\
&&\times x^3 \left\{e^{2 x} x^4 (2 \mu a+x) \left[8 (\mu a)^2 \left(2 x^2+x-1\right) \right.\right.  \nonumber \\ 
&&+ \left. \left. 4 \mu a x
   (4 x (x+1)-3)
+x^2 (4 x (2 x+3)-5)\right] \right.\nonumber \\ 
&&+  \left. e^{4 x} x^2 (2 \mu a+x) \left[32
   (\mu a)^4 (x-4) (2 x-1)
\right. \right.\nonumber\\
&&
+ 8 (\mu a)^3 x (4 x (4 x-15)+23)  \nonumber \\ 
&&+ \left.  4 (\mu a)^2 x^2 \left(28 x^2-94
   x+29\right) \right. \nonumber\\
&&
+ \left. 2 \mu a x^3 (8 x (3 x-8)+19)+x^4 (4 x (2 x-3)+7)\right] \nonumber\\
&&- \left. e^{6
   x} (2 \mu a+x)^3 
\left[-32 (\mu a)^4-16 (\mu a)^3 x-12 (\mu a)^2 x^2 \right. \right. \nonumber \\ 
&&+ \left. \left. 12 \mu a x^3+3 x^4\right]+x^7 \right\}.
\end{eqnarray}

\begin{figure}
\centering
\includegraphics[width=8cm, angle=0]{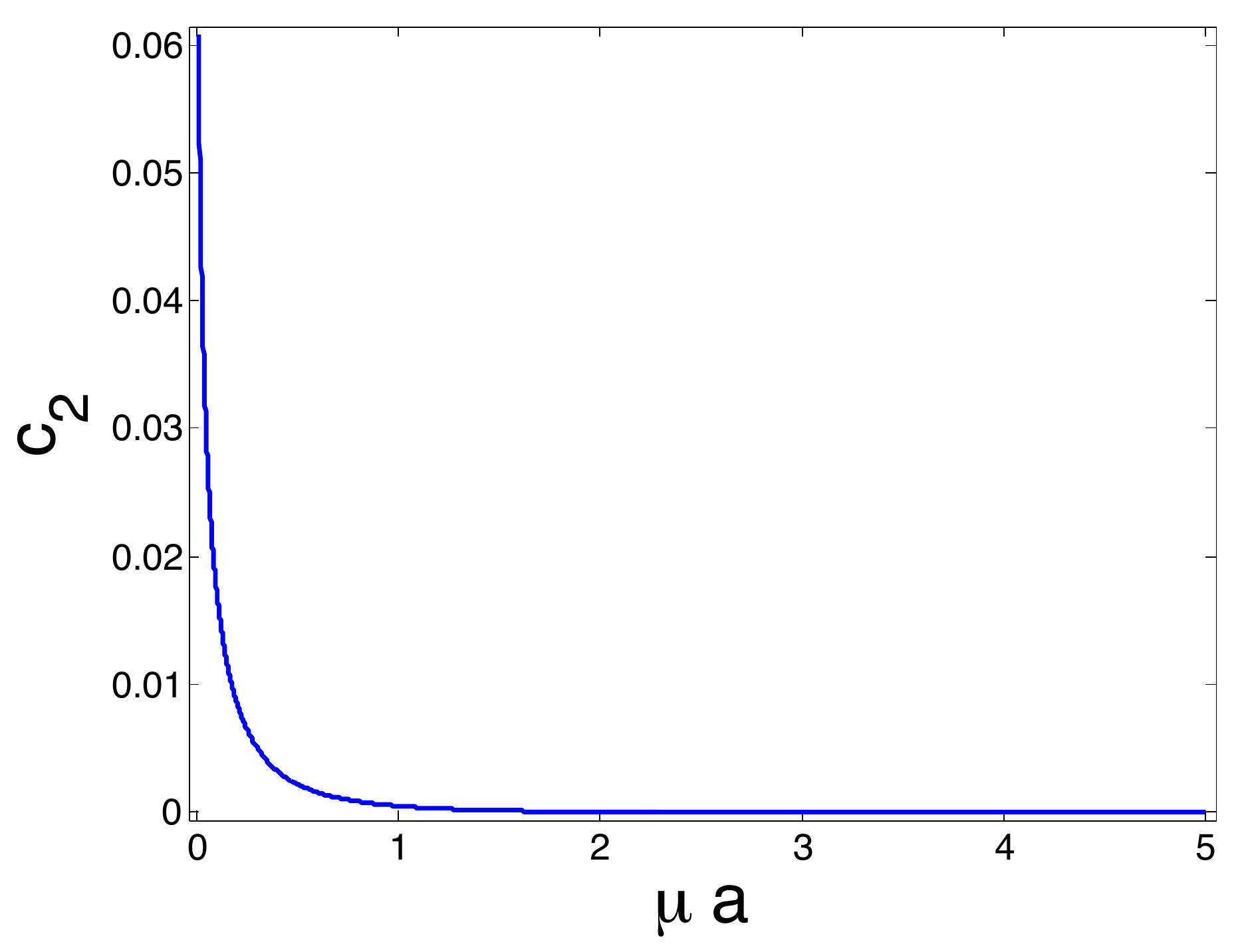}
\caption{Coefficient $c_{2}$ as a function of $\mu a$.} \label{fig2}
\end{figure}

 We have analyzed the behaviour of $c_2=c_{2l}+c_{2nl}$ both numerically
and analytically. As expected, $c_2$ vanishes as $\mu a\to \infty$, which
corresponds to the absence of mirrors. The most interesting limit is $\mu a
\ll 1$, since this case corresponds to almost-perfect Neumann boundary
conditions, a situation where the problems inherent to this case should
start to manifest themselves. Indeed: to begin with, one can readily check
that the integrand of Eq.(\ref{c2nl}) goes like $\sim 1/x$ for $\mu a \sim
0$, signalling the emergence of an infrared divergence, as anticipated in our previous work.
Moreover,  it can be shown analytically that, when $\mu a\ll 1$, 
 \begin{equation}\label{c2approx}
 c_2(\mu a)\simeq c_{2nl}(\mu a)\simeq -\frac{1}{16\pi}\log(\mu a)\, .
 \end{equation}
We have also checked this result by computing numerically the logarithmic
derivative of $c_2$ with respect to $\mu a$, which indeed tends to $-1/(16\pi)$
in the small-$\mu a$ limit. These results are illustrated in Figs. 2 and 3.
Fig. 2  depicts $c_{2}$ as a function of $\mu a$, showing that it vanishes for
large $\mu a$, while diverges in the opposite limit.  As a quantitative check for the small-$\mu a$ behavior
given in Eq.(\ref{c2approx}), 
in Fig. 3 we show the plot of the logarithmic derivative $\mu a\, dc_{2nl}/d\mu a$.

\begin{figure}
\centering
\includegraphics[width=8cm]{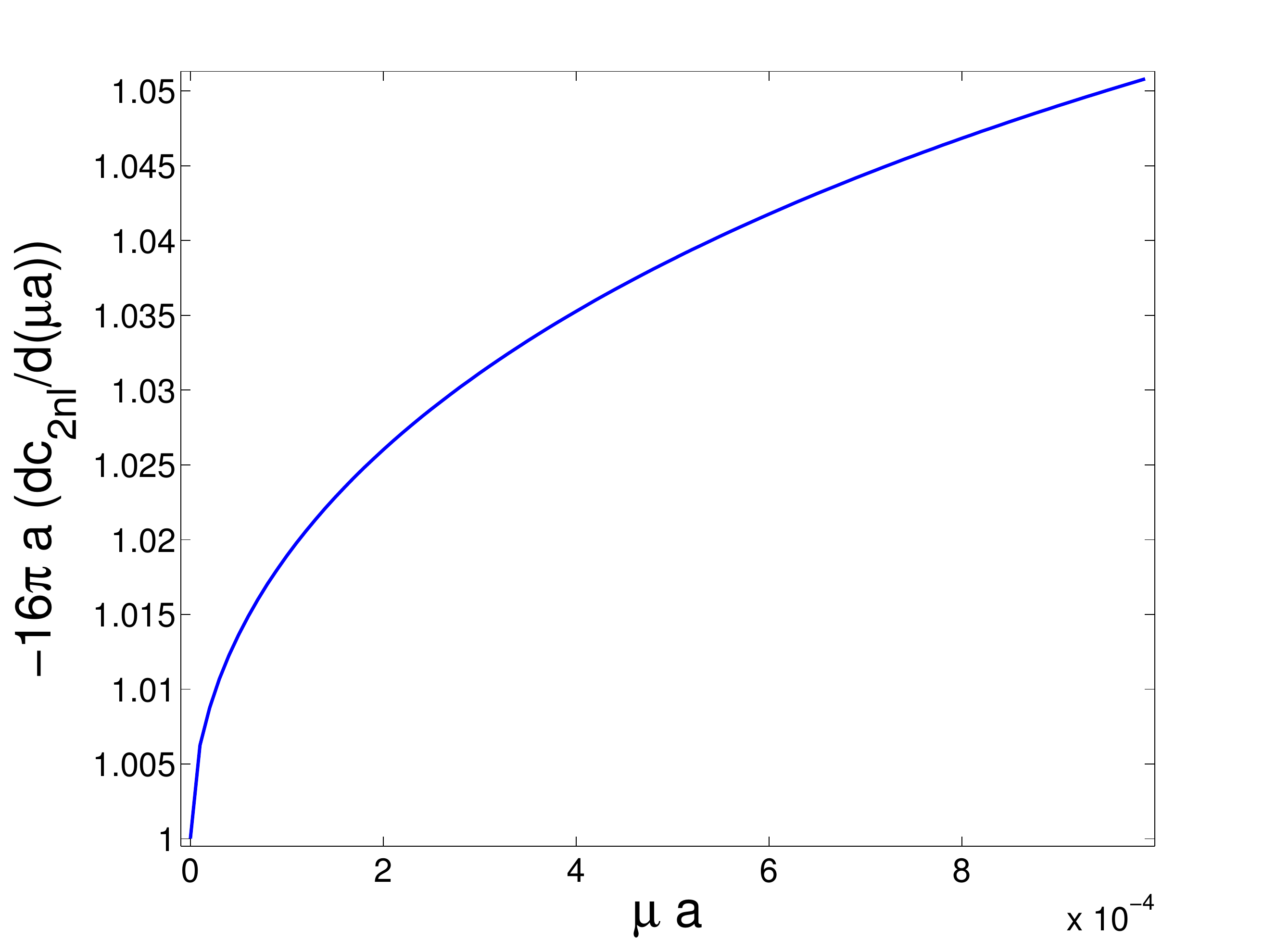}
\caption{Logarithmic derivative $-16 \pi\mu a\, dc_{2nl}/d\mu a$ as a function of $\mu a$.} \label{fig3}
\end{figure}

Collecting the results of this section, we can say that (up to the second
order) DE approximation to the Casimir energy reads, for small $\mu a$ (i.e., close to the perfect
case):
\begin{eqnarray}\label{eq:almostneumanm}
E_{\rm DE}[\psi] \;=\; &-&\frac{1}{16 \pi}\int_{-\infty}^\infty dx_1 \, \frac {1}{\psi(x_1)^2}
\left[ \zeta(3)\right. \nonumber \\
&+& \left.  \log[\mu\psi(x_1)]   \left(\frac{d\psi(x_1)}{dx_1} \right)^2  \right] \;,
\end{eqnarray} 
and this constitutes one of our main results. The first term is the PFA for the Casimir energy ($E_{\rm PFA})$.
The second term contains the first nontrivial correction to PFA ($E_{\rm NTLO}$) for an arbitrary boundary defined
by $\psi$. This equation shows that, although the DE is ill defined for Neumann
boundary conditions en $2+1$ dimensions, the non-analiticities in the DE
appear only in the case of perfect boundary conditions, that is, the
parameter $\mu$ acts as an infrared regulator.  The physical interpretation
anticipated in Ref.\cite{ded} for the appearance of non-analiticities for
Neumann boundary conditions is confirmed by this calculation. Indeed, for
$\mu=0$ the field contain massless modes, that are not present for $\mu\neq
0$.  
Besides, note that the relation between the Casimir energies computed for
different values of $\mu$ is encoded in the simple differential equation
\begin{equation}\label{rge}
\mu\frac{dE_{\rm DE}}{d\mu}\;=\; -\frac{1}{16 \pi}\int_{-\infty}^\infty dx_1 \, \frac {1}{\psi(x_1)^2}
\Big[ \zeta(3)\,+\,   \left(\frac{d\psi(x_1)}{dx_1} \right)^2  \Big] \;,
\end{equation} 
under the assumption that: $\mu\psi\ll 1$.
 
  \section{Examples}\label{sec:examp}
  
Let us consider a function $\psi$ describing a parabolic boundary
$\psi(x_1) = a + \frac{1}{2} x_1^2/R$ facing a straight line,
and approximate ($\mu \neq 0$) boundary conditions (EM or Neumann,
depending on the field considered).

Note that $a$ plays the role of the minimum distance between the two boundaries, while $R$ is the
curvature radius of the parabola at its vertex.
From Eq.(\ref{eq:almostneumanm}), we obtain  the DE
approximation to the Casimir energy, expected to be reliable, in this
example, when $\epsilon\equiv a/R\ll 1$.

The zeroth-order (PFA) term, calculated from the first line in Eq.
(\ref{eq:almostneumanm}), reads
\begin{equation}
E_{\rm PFA} = - \frac{\zeta(3)}{16\pi R}
\int_{-\infty}^{+\infty}\frac{du}{\left(\epsilon + \frac{u^2}{2}\right)^2}
= -\frac{\zeta(3)}{16\sqrt{2}R}\frac{1}{\epsilon^{\frac{3}{2}}}.
\label{pfaparabol}
\end{equation}

The NTLO correction comes from the second term in
Eq.(\ref{eq:almostneumanm}). To evaluate the integral, we
change the integration variable to $\psi(x_1) = z a$, so that:
\begin{eqnarray} 
E_{\rm NTLO} &=& - \frac{1}{8\pi R}\sqrt{\frac{2}{\epsilon}}
\int_{1}^{+\infty} dz \frac{\sqrt{z - 1}}{z^2} \log\left(\mu R \epsilon z
\right)\nonumber\\
&=&  - \frac{1}{16 R} \sqrt{\frac{2}{\epsilon}} \left[ 1 + \log(4 \mu R \epsilon)\right].
\label{neuparabol}
\end{eqnarray}
As expected, the NTLO correction to the DE expansion diverges
logarithmically in the limit of  perfect ($\mu=0$) Neumann boundary
conditions, but is finite in the imperfect case.

Besides, we have found it noteworthy that, fixing the value of $\mu$ and
taking the limit $\epsilon\to 0$, the ratio between the NTLO and  PFA terms
becomes independent of $\mu$ and reads
\begin{equation}
 \frac{E_{\rm NTLO}}{E_{\rm PFA} } \simeq \frac{2}{\zeta(3)} \epsilon \log\epsilon\, .
\end{equation}
This $\epsilon\log\epsilon$ behavior can be contrasted
with the case of perfect Dirichlet boundary conditions; in Ref.\cite{ded} we
have shown that, in $2+1$ dimensions,
\begin{eqnarray}\label{dirichletpfa}
E_{\rm DE}^{(D)}[\psi] \;=\; &-&\frac{1}{16 \pi}\int_{-\infty}^\infty dx_1 \, \frac {1}{\psi(x_1)^2}
\left[ \zeta(3)\right. \nonumber \\
&+& \left.   \frac{[1+6\zeta(3)]}{12}  \left(\frac{d\psi(x_1)}{dx_1} \right)^2  \right] 
\;,
\end{eqnarray} 
where the upper $(D)$ denotes Dirichlet boundary conditions. Computing
explicitly the integrals for the same example, we find
\begin{equation}
\frac{E_{\rm NTLO}^{(D)}}{E_{\rm PFA}^{(D)}} = \frac{\left(1 + 6 \zeta(3)\right)}{24 \zeta(3)}  \epsilon,\label{comparoD}
\end{equation}
which is linear in $\epsilon$,  in contrast with the quasi-perfect Neumann
case, that involves a logarithm.

\begin{figure}
\centering
\includegraphics[width=8cm]{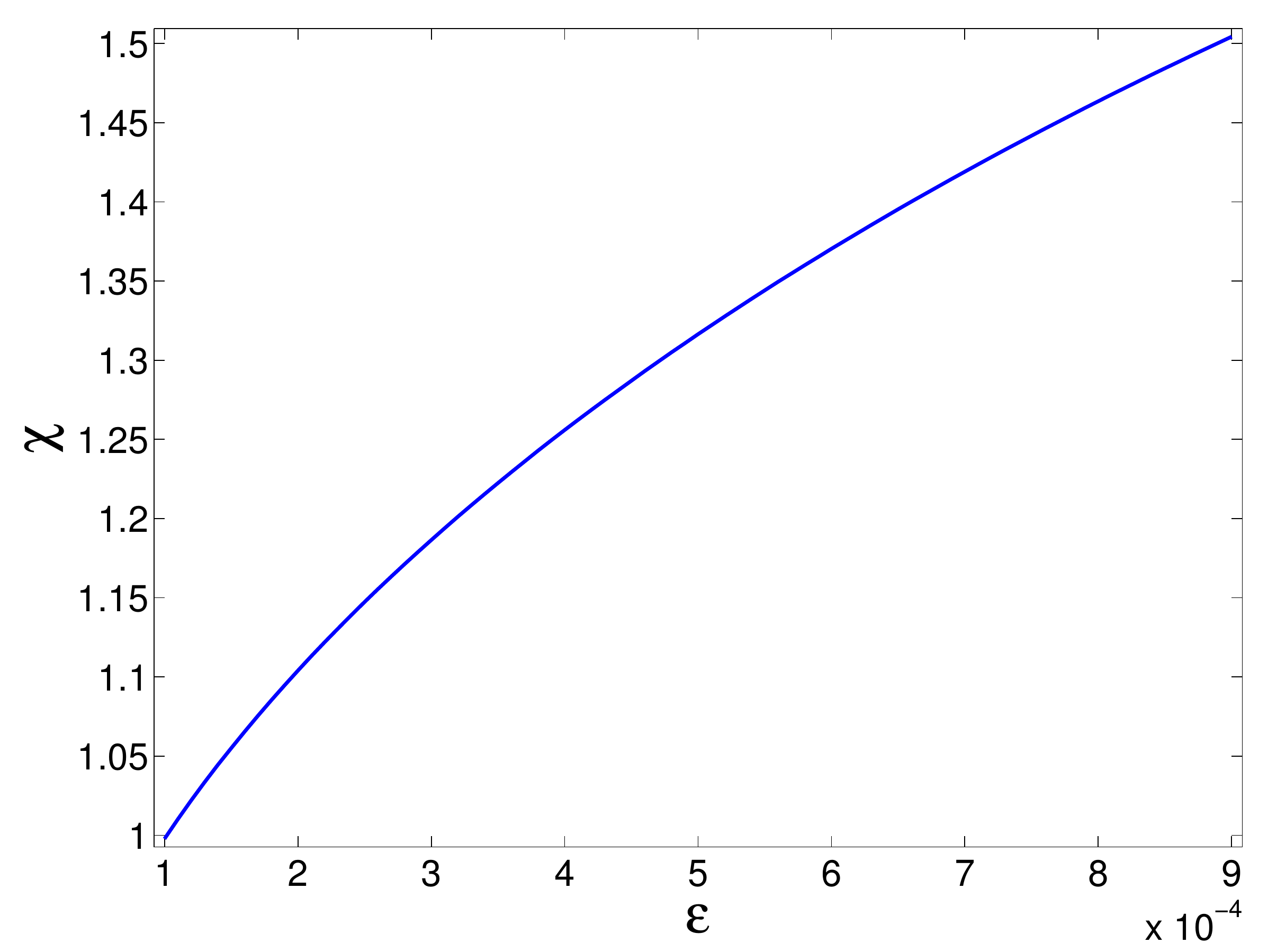}
\caption{$\chi$ (\ref{eq:defchi}) as a function of
$\epsilon=a/R$ (circle-line geometry) for $\mu R = 0.1$.
$\chi$ goes to 1 for small values of $\epsilon$, showing
that the NTLO correction in this case is similar to the one of the
parabola-line configuration.} \label{fig4}
\end{figure}

Let us now consider another example, that allows us to make contact with
previous results in the literature~\cite{Bordag2006}:  a circle in front a
straight line, which is the dimensionally reduced version of the
cylinder-plane case. 
Here, the function $\psi$ defining the curved contour is given
by $\psi(x_1) = a + R \left(1 - \sqrt{1 - x_1^2/R^2}\right)$, where $R$ is
the radius of the cylinder and $a$ is, again, the minimum distance between
the circle and the straight line.  On physical grounds, we expect  the
results for either a circle or a parabola in front of a line to be very
similar in the $\epsilon\ll 1$ limit. 
To check this assertion, since the integrals in  Eq. (\ref{eq:almostneumanm}) cannot
be computed analytically for the circle, we have performed a numerical
evaluation; in Fig.\ref{fig4} we plot the ratio:
\begin{equation}\label{eq:defchi}
\chi = \frac{ \frac{E_{\rm NTLO}}{E_{\rm PFA} } }{\frac{2}{\zeta(3)} \epsilon (1+ \log[4\mu R\epsilon])},
\end{equation}
which compares the ratio between NTLO and PFA terms to the value it should
have for the parabolic case, as a function of $\epsilon$. 
As expected, $\chi \to 1$ for small values of $\epsilon$; exhibiting a similar behavior to the 
parabola-line case.  
Therefore, we conclude that the NTLO correction to PFA is proportional to
$\epsilon\log\epsilon$ also for the circle-line geometry, a result that was
conjectured for perfect Neumann boundary conditions in
Ref.\cite{Bordag2006}.
  
\section{Conclusions}\label{sec:conc}

In this work, we present results which, we believe, may shed light on some
of the properties of the DE approximation to the Casimir energy. In
particular, we have studied a phenomenon pointed out in~\cite{ded},
namely, that for {\em perfect\/} Neumann boundary conditions in $2+1$ dimensions, 
the NTLO correction to the PFA cannot be written as the integral of a local term 
involving $\psi$ and up to two derivatives of that function. 
However, as the calculations presented here show explicitly, the NTLO
is perfectly well-defined and local when the mirrors are imperfect. 
In other words, the non-analyticity is an artifact of the idealization of the boundary
conditions. And this is corrected by an imperfection, no matter how tiny. 

It is worth emphasizing that perhaps a taming of the non
analyticity might also be obtained by introducing other, cruder infrared
cutoffs, like a mass term for the field. However, as we have shown, it is
sufficient to include a rather mild and  physically justified modification into
the game, which consists of an imperfection in the Neumann conditions,
parametrized by a constant that can be tuned to vary the mirrors'
properties.  

The same problem can be seen to appear when considering the high
temperature limit (for Neumann conditions) in $3+1$ dimensions~\cite{ded}.
Mathematically, the integral in momentum space that defines $Z(\psi)$ has
an infrared divergence. Based on analogies with results in the context of 
quantum field theory in non trivial backgrounds, we argued previously that 
the physical reason for the emergence of nonlocal corrections is the 
existence of gapless modes, which happens only for Neumann boundary conditions. 
We have also argued there that, were that the case, for imperfect (and therefore
more realistic) boundary conditions, the problem should be cured. We have
shown here that this is indeed the case, by providing an explicit example.

We considered the case example of the EM field in $2+1$
dimensions but, in what may be considered a by-product of our study, we have
seen that it is dual to a real scalar field, in the understanding that
perfect or imperfect conductor boundary conditions for the electromagnetic
field  correspond, respectively, to perfect of imperfect  Neumann boundary
conditions for the scalar field. 

Regarding  explicit results and examples, we have obtained the coefficients of the
second order DE, depending on a
parameter $\mu$ which measures the departure from perfect boundary
conditions, and applied them to evaluate the Casimir energy for the case 
of a parabola and a circle in front of a line. This  
enabled us to pinpoint the effect of the would-be infrared dominant 
contribution to the DE on the resulting energy, for specific geometries. We have also 
compared the results with those corresponding to Dirichlet boundary conditions,
where the problem of non-analyticity of the NTLO correction is not present.
 
\section*{Acknowledgements}
This work was supported by ANPCyT, CONICET, UBA and UNCuyo.
\newpage
\appendix
\section{ Intermediate results on the perturbative expansion for
$\Gamma$} \label{A}

We present here some technical details and intermediate results
corresponding to the calculation of the effective action to the second order in the function
$\eta$. We assume that $\psi(x_\parallel) = a + \eta(x_\parallel)$, with
$a$ equalling the average of $\psi$.

The term of order $1$ vanishes, and the others can be written in terms of
the expanded matrices ${\mathbb K}$, of elements $K_{\rm AB}$ (A, B = L,
R), which have  expressions that we present now. The zeroth order one
\begin{eqnarray}
{\mathbb K}^{(0)}(x_\parallel,x'_\parallel) &=& 
{\mathbb K}^{(0)}(x_\parallel - x'_\parallel) \nonumber \\ &=& \int
\frac{d^2k_\parallel}{(2\pi)^2} \, e^{i k_\parallel \cdot
(x_\parallel - x'_\parallel)} \; \widetilde{\mathbb K}^{(0)}(k_\parallel) \;, 
\end{eqnarray}
where
\begin{equation}
	\widetilde{\mathbb K}^{(0)}(k_\parallel) \;=\; 
	\frac{|k_\parallel| + 2\mu}{2}
	\left( 
		\begin{array}{cc}
			1 &  r(|k_\parallel|) e^{- |k_\parallel| a} \\
		       r(|k_\parallel|) e^{- |k_\parallel| a}   & 1
		\end{array}
	\right) \;,
\end{equation}
and $r(x) \;\equiv\; \frac{x}{x + 2 \mu}$. 

Regarding ${\mathbb K}^{(1)}$, we see that 
\begin{equation}
{\mathbb K}^{(1)}_{\rm RR}= {\mathbb K}^{(1)}_{\rm LL}= 0 \;,
\end{equation}
and the off-diagonal elements are given by:
\begin{eqnarray}
	&&K^{(1)}_{\rm LR}(x_\parallel,x'_\parallel) =  {\mathbb K}^{(1)}_{\rm RL}(x'_\parallel,x_\parallel) \nonumber \\
&&= -\frac{1}{2} 
\int \frac{d^2k_\parallel}{(2\pi)^2} \, e^{i k_\parallel \cdot (x_\parallel
- x'_\parallel)} \,\, e^{- |k_\parallel| a} \nonumber \\
&& \times \left[ |k_\parallel|^2 \eta(x_\parallel) + i  k^\alpha \partial_\alpha
\eta(x_\parallel)\right].
\end{eqnarray}
Finally, to the second order, ${\mathbb K}^{(2)}_{\rm LL}= 0$,  and:
\begin{eqnarray}
	&&{\mathbb K}^{(2)}_{\rm RR}(x_\parallel,x'_\parallel) = 
\int \frac{d^2k_\parallel}{(2\pi)^2} \, e^{i k_\parallel \cdot (x_\parallel
- x'_\parallel)} \nonumber \\
&& \times \left[ -\frac{1}{2} \big( |k_\parallel|^3 + \frac{k^\alpha k^\beta}{|k_\parallel|}
\partial_\alpha \partial'_\beta \big) \;\eta(x_\parallel)
\eta(x'_\parallel) \right. \nonumber \\ &&  - \left.  i  \,|k_\parallel| \eta(x_\parallel) (\partial_\alpha \eta(x_\parallel)
+ \partial'_\alpha \eta(x'_\parallel) ) \right] \;.
\end{eqnarray}
On the other hand, both ${\mathbb K}^{(2)}_{\rm LR}$ and ${\mathbb K}^{(2)}_{\rm LR}$  are non-vanishing, 
but it may be seen that they do not contribute to the second order term
under the assumption that the average of $\eta$ vanishes.

The Fourier transform of the inverse of ${\mathbb K}^{(0)}$ (we need it to
calculate the second order term) is given by:
\begin{eqnarray}
	& \big( &\widetilde{\mathbb K}^{(0)}\big)^{-1}(k_\parallel) = 
	\frac{2}{(|k_\parallel| + 2\mu)\left(1 - r^2(|k_\parallel|\right) e^{-2
|k_\parallel| a})} \nonumber \\
&\times&	\left( 
		\begin{array}{cc}
			1 & - r(|k_\parallel|) e^{- |k_\parallel| a} \\
		      - r(|k_\parallel|) e^{- |k_\parallel| a}   & 1
		\end{array}
	\right) .
\end{eqnarray}

The second order term, $\Gamma^{(2)}$ receives two contributions,
which we have denoted by $\Gamma^{(2,1)}$ and $\Gamma^{(2,2)}$ in
(\ref{eq:fbnn}). For each one of them we introduce a momentum kernel,
\begin{equation}
	\Gamma^{(2,a)}\,=\, \frac{1}{2} \int \frac{d^2k_\parallel}{(2\pi)^2}
	\,f^{(2,a)}(k_\parallel) \,
	|\tilde{\eta}(k_\parallel)|^2\;,\;\;a=1,2\;.
\end{equation}
An explicit evaluation of those two kernels yields:
\begin{eqnarray}
	& f^{(2,1)}(k_\parallel) \;=\; k_\parallel^2 \;\Big[\int
	\frac{d^2p_\parallel}{(2\pi)^2} \frac{r^2(|p_\parallel|)}{e^{2 a
	|p_\parallel|} - r^2(|p_\parallel|)} \,
\, \frac{\mu}{|p_\parallel| + 2 \mu} \Big]\nonumber\\
 &- \, \int \frac{d^2p_\parallel}{(2\pi)^2}
\frac{r^3(|p_\parallel|)}{e^{2 a |p_\parallel|} - r^2(|p_\parallel|)}\, 
\frac{[p_\parallel \cdot (p_\parallel + k_\parallel)]^2}{|p_\parallel|
|p_\parallel + k_\parallel|} \;,
\end{eqnarray}
and
\begin{widetext}
\begin{eqnarray}
	f^{(2,2)}(k_\parallel) &=& -  \int \frac{d^2p_\parallel}{(2\pi)^2}
	\left\{
	\frac{r^2(|p_\parallel|) r^2(|p_\parallel + k_\parallel|)}{\big(e^{2 a
		|p_\parallel|} - r^2(|p_\parallel|)\big) \big(e^{2 a
		|p_\parallel + k_\parallel|} -
r^2(|p_\parallel+k_\parallel|)\big)} \right. \nonumber \\ &+& \left. \frac{r(|p_\parallel|) r(|p_\parallel + k_\parallel|)}{\big(e^{2 a
		|p_\parallel|} - r^2(|p_\parallel|)\big) \big( 1 -
		r^2(|p_\parallel + k_\parallel|) e^{-2 a |p_\parallel +
k_\parallel|}\big)} \right\} \frac{[p_\parallel \cdot (p_\parallel + k_\parallel)]^2}{|p_\parallel|
|p_\parallel + k_\parallel|} .
\end{eqnarray}
\end{widetext}

\end{document}